\documentclass[A4paper]{article}
\usepackage{amsmath}
\usepackage{amssymb}
\usepackage{graphicx,psfrag,epsf}
\usepackage{enumerate}
\usepackage{natbib}
\usepackage{url} 
\usepackage{fullpage}
\usepackage{algorithm}
\usepackage{algorithmic}

\newcommand{\blind}{1}

\renewcommand\footnotemark{}

\begin{document}

\def\spacingset#1{\renewcommand{\baselinestretch}%
{#1}\small\normalsize} \spacingset{1}


\if1\blind
{
  \title{\bf \Large An approach for finding fully Bayesian optimal designs using normal-based approximations to loss functions
  \vspace{0cm}
  }
  \author{\large Antony M. Overstall\hspace{.2cm}\\
    \large Southampton Statistical Sciences Research Institute,\\ \large University of Southampton,\\ \large Southampton, UK\\ \large (A.M.Overstall@soton.ac.uk) \\[1ex]
    \large James M. McGree \& Christopher C. Drovandi \\
    \large School of Mathematical Sciences,\\ \large Queensland University of Technology,\\ \large Brisbane, Australia \\ \large (james.mcgree@qut.edu.au) \& (c.drovandi@qut.edu.au)}
    \date{\vspace{-1.2cm}}
  \maketitle
} \fi

\if0\blind
{
  \bigskip
  \bigskip
  \bigskip
  \begin{center}
    {\LARGE\bf Title}
\end{center}
  \medskip
} \fi

\bigskip
\begin{abstract}
The generation of decision-theoretic Bayesian optimal designs is complicated by the significant computational challenge of minimising an analytically intractable expected loss function over a, potentially, high-dimensional design space. A new general approach for approximately finding Bayesian optimal designs is proposed which uses computationally efficient normal-based approximations to posterior summaries to aid in approximating the expected loss. This new approach is demonstrated on illustrative, yet challenging, examples including hierarchical models for blocked experiments, and experimental aims of parameter estimation and model discrimination. Where possible, the results of the proposed methodology are compared, both in terms of performance and computing time, to results from using computationally more expensive, but potentially more accurate, Monte Carlo approximations. Moreover the methodology is also applied to problems where the use of Monte Carlo approximations is computationally infeasible.

\end{abstract}

\noindent%

\section{Introduction}\label{sec:intro}

The process of designing a physical experiment fits naturally within the Bayesian approach to statistical inference. Prior information on parameters and models can be represented by prior distributions, and the experimental aim encapsulated in a decision-theoretic framework by the loss function. A Bayesian optimal design is found by minimising the expected loss function over the space of all possible designs, i.e. the design space, where the expectation is with respect to the joint distribution of all unknown quantities, i.e. parameters, models and experimental responses.

Formally, suppose the experiment consists of $n$ runs where the $i$th run (for $i=1,\dots,n$) involves measuring response $y_i$ having specified settings $\mathbf{d}_i = \left(d_{i1},\dots,d_{ik}\right)$ for the $k$ controllable factors. Let $\mathbf{d} = \left(\mathbf{d}_1,\dots,\mathbf{d}_n\right) \in \mathcal{D}$ be the $W \times 1$ vector giving the design where $W=nk$ and $\mathcal{D} \subset \mathbb{R}^W$ denotes the $W$-dimensional design space. Assume there is a set, $\mathcal{M}$, of competing statistical models. Model $m \in \mathcal{M}$ posits a probability distribution, $\mathrm{F}_m$, for $\mathbf{y}=\left(y_1,\dots,y_n\right)$ which is completely specified up to an unknown $p_m \times 1$ vector of parameters $\boldsymbol{\theta}_m$. Bayesian inference on models and/or parameters is based on their joint posterior distribution given by Bayes' theorem as
\begin{equation}
\pi \left(\boldsymbol{\theta}_m,m|\mathbf{y},\mathbf{d}\right) \propto  \pi \left(\mathbf{y}|\boldsymbol{\theta}_m,m,\mathbf{d}\right)\pi \left(\boldsymbol{\theta}_m|m\right)\pi(m),
\label{eqn:post}
\end{equation}
where $\pi \left(\mathbf{y}|\boldsymbol{\theta}_m,m,\mathbf{d}\right)$ is the probability mass/density function of $\mathrm{F}_m$; $\pi \left(\boldsymbol{\theta}_m|m\right)$ is the probability density function of the prior distribution of $\boldsymbol{\theta}_m$; and $\pi(m)$ is the prior model probability of model $m$. Note how $\pi \left(\mathbf{y}|\boldsymbol{\theta}_m,m,\mathbf{d}\right)$ depends on the design and this induces a dependence of the posterior on the design.

The aim of the experiment is represented by a loss function, which can be tailored to experimental aims of parameter estimation or model discrimination. In general, the loss function is given by $\lambda(\boldsymbol{\theta}_m,m,\mathbf{y},\mathbf{d})$. Essentially, it compares a summary of the posterior distribution \citep[e.g.][pgs 13-14]{ohagan_forster_2004} for $\boldsymbol{\theta}_m$ and $m$ (conditional on $\mathbf{y}$ and $\mathbf{d}$) to the true values of $\boldsymbol{\theta}_m$ and $m$. However, $\boldsymbol{\theta}_m$, $m$ and $\mathbf{y}$ are unknown so a Bayesian optimal design \cite[e.g.][]{ChalonerVerdinelli} is found by minimising the expected loss
\begin{equation}
L(\mathbf{d}) = \mathrm{E}_{\boldsymbol{\theta}_m,m,\mathbf{y}|\mathbf{d}} \left[ \lambda(\boldsymbol{\theta}_m,m,\mathbf{y},\mathbf{d}) \right]
\label{eqn:exploss}
\end{equation}
over the design space, $\mathcal{D}$, where the expectation in (\ref{eqn:exploss}) is with respect to the joint distribution of $\boldsymbol{\theta}_m$, $m$ and $\mathbf{y}$. 

\citet[Section 2.1]{robert_2007} discusses how Bayesian inference relies on the specification of three components: a) the set of models, $\mathcal{M}$; b) the joint prior distribution, given by $\pi(\boldsymbol{\theta}_m,m)$; and c) the loss function, $\lambda(\boldsymbol{\theta}_m,m,\mathbf{y},\mathbf{d})$. Notwithstanding the difficulties faced in specifying these three components \citep[e.g.][Chapter 6]{ohagan_forster_2004}, once in place, a Bayesian optimal design is conceptually straightforward to define. However, finding such a design in practice is hindered by the computational challenge of minimising the expected loss. Typically, the expected loss is given by a multi-dimensional analytically intractable integral, i.e. as given by (\ref{eqn:exploss}). In the two decades since the seminal review of Bayesian design by \cite{ChalonerVerdinelli}, there have been few general-purpose approaches to finding such designs, as recently highlighted by \cite{ryan_drovandi_mcgree_pettitt_2015} and \cite{woods_ace_2016}. Note that being able to find the exact optimal design for an arbitrary problem is, at present, an unrealistic goal. Instead, the aim is to find a design \textquotedblleft close" to the optimal design, termed a near-optimal design by \cite{hamada_etal_2001}.

Existing approaches to approximately finding Bayesian designs can be divided into two broad strategies. First, the simulation-based approach of \cite{muller_1999} arranges a joint distribution for $\boldsymbol{\theta}_m$, $m$, $\mathbf{y}$ and design $\mathbf{d}$ given by
\begin{equation}
h(\boldsymbol{\theta}_m,m,\mathbf{y},\mathbf{d}) \propto \left(c - \lambda(\boldsymbol{\theta}_m,m,\mathbf{y},\mathbf{d})\right) \pi \left(\mathbf{y}|\boldsymbol{\theta}_m,m,\mathbf{d}\right)\pi \left(\boldsymbol{\theta}_m|m\right)\pi(m),
\label{eqn:h}
\end{equation}
where $c \ge \sup \lambda(\boldsymbol{\theta}_m,m,\mathbf{y},\mathbf{d})$. The marginal mode of $\mathbf{d}$ corresponds to the Bayesian optimal design. The so-called M\"{u}ller algorithm essentially proceeds by using simulation methods to generate a sample from the joint distribution of $\boldsymbol{\theta}_m$, $m$, $\mathbf{y}$ and $\mathbf{d}$ and to use this sample to estimate the marginal mode of $\mathbf{d}$. This approach has been further modified by \cite{muller_sanso_iorio_2004} and \cite{amzal_bois_parent_robert_2006}. However, due to the difficulty in implementing efficient sampling methods, the M\"{u}ller algorithm is difficult to implement for high dimensional design spaces with the limit typically considered to be just $W=4$ \citep[e.g.][]{ryan_drovandi_mcgree_pettitt_2015}. 

Alternatively, the second broad strategy is the smoothing-based approach reliant on the following Monte Carlo approximation to the expected loss
\begin{equation}
\hat{L}(\mathbf{d}) = \frac{1}{B} \sum_{b=1}^B \lambda(\boldsymbol{\theta}^b_m,m^b,\mathbf{y}^b,\mathbf{d}),
\label{eqn:mcapprox}
\end{equation}
where $\left\{\boldsymbol{\theta}^b_m,m^b,\mathbf{y}^b\right\}_{b=1}^B$ is a sample generated from the joint distribution $\boldsymbol{\theta}_m$, $m$ and $\mathbf{y}$. The stochastic nature of Monte Carlo approximation means that $\hat{L}(\mathbf{d})$ is not a smooth function and makes application of standard optimisation methods (including heuristic methods) difficult. Instead, \cite{MullerParmigiani} proposed an approach whereby $\hat{L}(\mathbf{d})$ is evaluated at a series of designs. A statistical model (or smoother) is then fitted which builds a relationship between design and expected loss allowing prediction of the expected loss for any design. This prediction is minimised in place of the true expected loss to give an approximation to the Bayesian optimal design. Similar to the simulation-based M\"{u}ller algorithm, the scalability of this approach to higher dimensional design spaces remains an issue. The chosen smoother needs to balance the increased flexibility required for adequate predictive accuracy with the computational effort of the increased number of evaluations of $\hat{L}(\mathbf{d})$ required. \cite{MullerParmigiani} employed local regression models and considered design spaces up to $W=2$. More recently, \cite{Weaver} used a Gaussian process model \citep[e.g.][]{santner_2003} and considered an application with $W=3$ and \cite{Jones} used Bayes linear analysis \citep{GoldsteinWooff} and considered up to $W=9$. To increase the applicability to higher dimensional design spaces, \cite{woods_overstall_2016} proposed the approximate coordinate exchange (ACE) algorithm whereby a cyclic descent algorithm (commonly called coordinate exchange in the design of experiments literature, \citealt{meyer_nachtsheim_1995}) is used to minimise the expected loss. Very briefly, a Gaussian process prediction of the expected loss is sequentially minimised over each one-dimensional element of the design space. This can be seen as a generalisation of the approaches of \cite{MullerParmigiani} and \cite{Weaver} to higher dimensional design spaces via the use of coordinate exchange, and thus allowed consideration of examples with design spaces of dimensionality nearly two orders of magnitude greater than previously addressed in the literature.

Both the M\"{u}ller algorithm and the smoothing-based approaches require evaluations of the loss function $\lambda(\boldsymbol{\theta}_m,m,\mathbf{y},\mathbf{d})$ either in the evaluation of $h(\boldsymbol{\theta}_m,m,\mathbf{y},\mathbf{d})$ or $\hat{L}(\mathbf{d})$, respectively. In both cases, a large number of evaluations of the loss function will be required to find a design. Since the loss function depends on $\mathbf{y}$ through the posterior distribution of $\boldsymbol{\theta}_m$ and $m$, and given that this distribution will typically be analytically intractable, a further approximation is required. The most obvious approach to this problem is to use an additional Monte Carlo approximation, the exact nature of which depends on the chosen loss function. In the case of the Monte Carlo approximation to the expected loss, $\hat{L}(\mathbf{d})$, given by (\ref{eqn:mcapprox}), this will result in a nested or double loop Monte Carlo (DLMC) approximation to the expected loss where the inner loop refers to the approximation to the loss function and the outer loop to the approximation to the expected loss. Let $\tilde{B}$ denote the Monte Carlo sample size in the inner loop with $B$ being the corresponding value in the outer loop. For typical loss functions, DLMC can induce a bias of order $\tilde{B}^{-1}$ in the approximation \citep[e.g.][]{Ryan_2003,rainforth_et_al_2016} to the expected loss. Moreover, the computational complexity of this approach is typically in the order of $B \times \tilde{B}$ evaluations of $\pi(\mathbf{y}|\boldsymbol{\theta}_m,m,\mathbf{d})$ for each $m \in \mathcal{M}$. Since $B$ and $\tilde{B}$ will typically be $\mathcal{O}(10^3)$ or higher, this will be a computationally expensive approach. The result of which is that, even by using the ACE algorithm, finding Bayesian optimal designs with the DLMC approximation to the expected loss has been confined to simple problems where the number of models under consideration is $|M|=1$ and inference has been focused on parameter estimation.

In this paper, we consider using normal-based approximations to the posterior distribution of $\boldsymbol{\theta}_m$, centred around the posterior mode of $\boldsymbol{\theta}_m$, and how these can be used to approximate, in principle, any loss function. The normal-based Laplace approximation has previously been used to approximate the commonly-used self-information loss function (see Section~\ref{sec:losses}) for a non-linear model by \cite{laplace} for a low-dimensional design space and under no model uncertainty. However application to other loss functions has not previously been considered. We apply the new methodology to illustrative examples which are challenging in the context of Bayesian optimal design. In cases where the computationally more expensive DLMC approach is feasible, we show, empirically, that the difference in performance (measured in terms of expected loss) between designs found under the two different approximations is negligible. We also apply the proposed approach to problems where use of the DLMC approximation to find a design would be computationally infeasible.

\section{Methodology}\label{sec:meth}

\subsection{Normal-based approximations to posterior quantities}\label{sec:normbase}

As discussed in Section~\ref{sec:intro}, a typical loss function compares a summary of the joint posterior distribution of $\boldsymbol{\theta}_m$ and $m$ to the \textquotedblleft true" values of $\boldsymbol{\theta}_m$ and $m$ in a way that is relevant to the experimental aim. However, the joint posterior distribution is usually not available in closed form. The methodology in this paper is based on forming an approximation to this joint distribution using normal-based approximations and therefore providing an alternative approximation to the loss function than the Monte Carlo approximation. The normal-based approximation to the loss function, denoted by $\tilde{\lambda}(\boldsymbol{\theta}_m,m,\mathbf{y},\mathbf{d})$, can then be substituted into the Monte Carlo approximation to the expected loss, given by (\ref{eqn:mcapprox}) and we refer to such an approximation as normal-based Monte Carlo (NBMC). We can then use the ACE algorithm or one of the other smoothing based approaches discussed in Section~\ref{sec:intro}. Conversely, we may also substitute $\tilde{\lambda}(\boldsymbol{\theta}_m,m,\mathbf{y},\mathbf{d})$ into the density of the joint distribution over $\boldsymbol{\theta}_m$, $m$, $\mathbf{y}$ and $\mathbf{d}$, given by (\ref{eqn:h}), and apply the M\"{u}ller algorithm.

First, note that the joint posterior distribution of $\boldsymbol{\theta}_m$ and $m$, given by (\ref{eqn:post}), can be decomposed as follows \citep[e.g.][page 169]{ohagan_forster_2004}
$$\pi \left(\boldsymbol{\theta}_m,m | \mathbf{y},\mathbf{d}\right) = \pi(\boldsymbol{\theta}_m | \mathbf{y},m,\mathbf{d})\pi(m|\mathbf{y},\mathbf{d}),$$
where the posterior model probability of model $m \in \mathcal{M}$ is given by
\begin{equation}
\pi(m|\mathbf{y},\mathbf{d}) = \frac{\pi(\mathbf{y}|m,\mathbf{d}) \pi(m)}{\sum_{m \in \mathcal{M}} \pi(\mathbf{y}|m,\mathbf{d}) \pi(m)},
\label{eqn:pmp}
\end{equation}
and the posterior distribution of $\boldsymbol{\theta}_m$ (conditional on model $m \in \mathcal{M}$) is given by
\begin{equation}
\pi \left(\boldsymbol{\theta}_m|\mathbf{y},m,\mathbf{d}\right) = \frac{\pi\left(\mathbf{y}|\boldsymbol{\theta}_m,m,\mathbf{d}\right)\pi \left(\boldsymbol{\theta}_m|m\right)}{\pi(\mathbf{y}|m,\mathbf{d})},
\label{eqn:posttheta}
\end{equation}
with
\begin{equation}
\pi(\mathbf{y}|m,\mathbf{d}) = \int_{\Theta_m} \pi\left(\mathbf{y}|\boldsymbol{\theta}_m,m,\mathbf{d}\right)\pi \left(\boldsymbol{\theta}_m|m\right) \mathrm{d}\boldsymbol{\theta}_m,
\label{eqn:marglik}
\end{equation}
usually called the marginal likelihood or evidence \citep{frielwyse} for model $m \in \mathcal{M}$.

The posterior model probabilities are completely determined by the marginal likelihoods. Therefore for cases where there is model uncertainty, i.e. $|\mathcal{M}|>1$, it will be necessary to approximate $\pi(\mathbf{y}|m,\mathbf{d})$ for all $m \in \mathcal{M}$. First, let the posterior mode for model $m \in \mathcal{M}$ be denoted by $\hat{\boldsymbol{\theta}}_m(\mathbf{y}) = \arg \max_{\boldsymbol{\theta}_m \in \Theta_m} \pi(\mathbf{y}|\boldsymbol{\theta}_m,m,\mathbf{d})\pi(\boldsymbol{\theta}_m|m)$.  Now let
$$\hat{\Sigma}_m(\mathbf{y}) = \mathbf{H}(\hat{\boldsymbol{\theta}}_m(\mathbf{y});m)^{-1},$$
where 
$$\mathbf{H}(\boldsymbol{\theta}_m;m) = \mathrm{E}_{\mathbf{y}|\boldsymbol{\theta}_m,m} \left[ \frac{\partial \log \pi (\mathbf{y}|\boldsymbol{\theta}_m,m,\mathbf{d})}{\partial \boldsymbol{\theta}_m} \frac{\partial \log \pi (\mathbf{y}|\boldsymbol{\theta}_m,m,\mathbf{d})}{\partial \boldsymbol{\theta}_m^T}\right] - \frac{\partial^2 \log \pi(\boldsymbol{\theta}_m|m)}{\partial \boldsymbol{\theta}_m \partial \boldsymbol{\theta}_m^T},$$
i.e. the Fisher information matrix minus the second derivative of the log prior density. A second-order Taylor series expansion of the log integrand in (\ref{eqn:marglik}) about the posterior mode yields the following so-called Laplace approximation \citep[e.g.][]{gelman_etal_2014,laplace} to the marginal likelihood for model $m \in \mathcal{M}$
\begin{equation}
\tilde{\pi}(\mathbf{y}|m,\mathbf{d}) = \left(2 \pi\right)^{\frac{p_m}{2}} |\hat{\Sigma}_m(\mathbf{y})|^{\frac{1}{2}} \pi(\mathbf{y}|\hat{\boldsymbol{\theta}}_m(\mathbf{y}),m,\mathbf{d})\pi(\hat{\boldsymbol{\theta}}_m(\mathbf{y})|m,\mathbf{d}).
\label{eqn:7a}
\end{equation}
The posterior model probabilities are now approximated via
\begin{equation}
\tilde{\pi}(m|\mathbf{y},\mathbf{d}) = \frac{\tilde{\pi}(\mathbf{y}|m,\mathbf{d}) \pi(m)}{\sum_{m \in \mathcal{M}} \tilde{\pi}(\mathbf{y}|m,\mathbf{d}) \pi(m)}.
\label{eqn:7b}
\end{equation}

Furthermore, we approximate the posterior distribution of $\boldsymbol{\theta}_m$ (conditional on $m$), by a normal distribution with mean $\hat{\boldsymbol{\theta}}_m(\mathbf{y})$ and variance $\hat{\Sigma}_m(\mathbf{y})$, i.e.
\begin{equation}
\mathrm{N}\left(\hat{\boldsymbol{\theta}}_m(\mathbf{y}), \hat{\Sigma}_m(\mathbf{y}) \right).
\label{eqn:approxdist}
\end{equation}
The tractability of the normal distribution means there are now many direct approximations to posterior summaries of interest \citep[e.g.][page 237]{ohagan_forster_2004}. For example, trivially, the posterior median of $\boldsymbol{\theta}_m$ conditional on $m$ is approximated by the posterior mode. These approximations to posterior summaries of interest can then be used to approximate many loss functions of practical interest (see Section~\ref{sec:losses} for examples). 

Since the posterior distribution of $\boldsymbol{\theta}_m$ converges to a normal distribution as $n \to \infty$ \citep[e.g.][pages 585-588]{gelman_etal_2014} the approximation will be more accurate for large $n$. The approximation can be expected to be poor when the true posterior distribution of $\boldsymbol{\theta}_m$ is multi-modal or has significant skewness. In the latter case, the approximation could be improved by a reparameterisation \citep[e.g.][]{achcar_smith_1990}.

\subsection{Finding the posterior mode and Fisher information} \label{sec:mode}
The approximations above are reliant on finding the posterior mode, $\hat{\boldsymbol{\theta}}_m(\mathbf{y})$, for each $m \in \mathcal{M}$. This will need to be accomplished for each $\mathbf{y}^b$ in the sample $\left\{\boldsymbol{\theta}_{m^b}^b,m^b,\mathbf{y}^b\right\}_{b=1}^B$, from the joint distribution of $\boldsymbol{\theta}_m$, $m$ and $\mathbf{y}$, to evaluate the NBMC approximation to the expected loss. To do this we use a scoring algorithm \citep[e.g.][pgs 254-257]{lange_2013}, i.e. Newton's method where evaluation of the Hessian matrix of the log posterior density is replaced by evaluation of $-\mathbf{H}(\boldsymbol{\theta}_m;m)$. Specifically, let
$$\mathbf{f}(\boldsymbol{\theta}_m;m) = \frac{\partial \log \pi(\mathbf{y}|\boldsymbol{\theta}_m,m,\mathbf{d})}{\partial \boldsymbol{\theta}_m} + \frac{\partial \log \pi(\boldsymbol{\theta}_m|m)}{\partial \boldsymbol{\theta}_m},$$
be the gradient of the log posterior density with respect to $\boldsymbol{\theta}_m$.

For $r = 0,1,2,\dots$, the scoring algorithm iterates through the following steps
$$\boldsymbol{\theta}_m^{(r+1)} = \boldsymbol{\theta}_m^{(r)} + \kappa \mathbf{H}\left(\boldsymbol{\theta}_m^{(r)};m\right)^{-1}\mathbf{f}(\boldsymbol{\theta}_m^{(r)};m),$$
for some $0<\kappa \le 1$, until convergence. In the examples in this paper, we use $\kappa = \frac{1}{4}$ and a starting value of $\boldsymbol{\theta}_m^{(0)} =\mathrm{E}\left(\boldsymbol{\theta}_m|m\right)$, i.e. the prior mean. Convergence is deemed to have occurred when \newline
$\left(\boldsymbol{\theta}_m^{(r+1)} - \boldsymbol{\theta}_m^{(r)}\right)^T\left(\boldsymbol{\theta}_m^{(r+1)} - \boldsymbol{\theta}_m^{(r)}\right) < \epsilon$, where $\epsilon = 10^{-4}$.

\vspace{0.5cm}

The scoring algorithm requires the repeated inversion of the $p_m \times p_m$ matrix $\mathbf{H}(\boldsymbol{\theta}_m;m)$. Often experiments are conducted in blocks where a block consists of homogenous experimental units. A suitable model \citep[e.g.][Chapter 17]{pawitan_2013} in this case is a hierarchical (or mixed model) where the effect of the block is accounted for by using block-specific parameters (sometimes referred to as random effects) for each block, which are assumed to be independent having a common prior distribution. A consequence is that the number of parameters, $p_m$, for these types of models is proportional to the number of blocks and therefore can be large. However due to the conditional independence structure exhibited by hierarchical models, 
$\mathbf{H}(\boldsymbol{\theta}_m;m)$ will be sparse leading to computationally efficient methods for finding the inverse.

\section{Examples} \label{sec:examples}

In this section we begin by discussing a range of exemplar loss functions and how they can be approximated using the approach described in Section~\ref{sec:meth}. This selection of loss functions is not exhaustive but instead demonstrate how typical loss functions may be approximated by using the approximations outlined in Section~\ref{sec:meth}. We then apply the proposed methodology to find designs for experiments involving standard and hierarchical logistic regression (Section~\ref{sec:log_reg}) and a non-linear model (Section~\ref{sec:BH}), for experimental aims of parameter estimation and model discrimination. In the examples we use the ACE algorithm (briefly described in Section~\ref{sec:intro}) to find the optimal design since this is the only method in the literature suitable for finding Bayesian designs for the dimensionality of design space considered. However, the approximations to the loss function could be applied with any method such as the M\"{u}ller algorithm or another smoothing-based method. A more detailed description of the ACE algorithm is provided in Appendix \ref{app:ace} with a description of the choice of tuning parameters. 

All designs found in Sections~\ref{sec:log_reg} and~\ref{sec:BH} can be reproduced via the \texttt{R} \citep{RRR} package \texttt{NBdesigns} \citep{NBdesigns}. This is available as a Supplementary Material to this paper. This package allows users to compare future computational methodology to the approaches described in this paper via the benchmark examples considered.

\subsection{Loss functions} \label{sec:losses}

The loss functions considered in this section can be categorised into those for a) parameter estimation (Section~\ref{sec:paraest}); and b) model discrimination (Section~\ref{sec:moddisc}).

\subsubsection{Parameter estimation} \label{sec:paraest}

Suppose interest lies in the $q \times 1$ vector of transformed parameters $\boldsymbol{\phi} = g_m(\boldsymbol{\theta}_m)$, where $q \le \min_{m \in \mathcal{M}} p_m$, the $g_m$ are a set of one-to-one and invertible functions, and $\boldsymbol{\phi}$ has a consistent interpretation for all $m \in \mathcal{M}$. Inference about $\boldsymbol{\phi}$ is based on the model-averaged posterior distribution \citep[][pages 171-174]{ohagan_forster_2004} given by
$$\pi_{\phi}(\boldsymbol{\phi}|\mathbf{y},\mathbf{d}) = \sum_{m \in \mathcal{M}} \pi_{\phi}(\boldsymbol{\phi}|m,\mathbf{y},\mathbf{d}) \pi(m|\mathbf{y},\mathbf{d}),$$
where we have introduced a subscript of $\phi$ on the density function to indicate it refers to the transformed parameters. Consider the following loss functions representing the experimental aim of estimating $\boldsymbol{\phi}$. In each case, a special case occurs when there is no model uncertainty, i.e. $|\mathcal{M}|=1$, in which case we are interested in parameter estimation under a single model.

\paragraph{Self-information loss}
The self-information (SI) loss is
\begin{equation}
\lambda_{SI}(\boldsymbol{\phi}, \mathbf{y}, \mathbf{d}) = \log \pi_{\phi}(\boldsymbol{\phi}) - \log \pi_{\phi}(\boldsymbol{\phi}|\mathbf{y},\mathbf{d}).
\label{eqn:siloss}
\end{equation}
Minimising the expected SI loss is equivalent to maximising the expected Shannon information gain \citep{Lindley1956} and expected Kullback-Liebler divergence between prior and posterior distributions. Note that the expected SI loss is non-positive. Typically, the density of the posterior distribution of $\boldsymbol{\phi}$, given by $\pi_{\phi}(\boldsymbol{\phi}|\mathbf{y},\mathbf{d})$, in the SI loss will be analytically intractable. However, we can approximate it by first approximating the posterior distribution of $\boldsymbol{\theta}_m$ by (\ref{eqn:approxdist}) and then deriving the approximate distribution of $\boldsymbol{\phi} = g_m(\boldsymbol{\theta}_m)$ (conditional on $m$) after taking a first-order Taylor series expansion of $\boldsymbol{\phi} = g_m(\boldsymbol{\theta}_m)$ about $\hat{\boldsymbol{\theta}}_m(\mathbf{y})$ \citep[e.g.][Chapter 4]{khuri_2003}. This leads to the following approximation to $\pi_{\phi}(\boldsymbol{\phi}|\mathbf{y},\mathbf{d})$ 
\begin{equation}
\tilde{\pi}_{\phi}(\boldsymbol{\phi}|\mathbf{y},\mathbf{d}) = \sum_{m \in \mathcal{M}} \tilde{\pi}_{\phi}(\boldsymbol{\phi}|m,\mathbf{y},\mathbf{d}) \tilde{\pi}(m|\mathbf{y},\mathbf{d}),
\label{eqn:star}
\end{equation}
where $\tilde{\pi}(m|\mathbf{y},\mathbf{d})$ is given by (\ref{eqn:7b}) and $\tilde{\pi}_{\phi}(\boldsymbol{\phi}|m,\mathbf{y},\mathbf{d})$ is the density of
\begin{equation}
\mathrm{N}\left( g_m(\hat{\boldsymbol{\theta}}_m(\mathbf{y})), \left. \frac{\partial g_m(\boldsymbol{\theta}_m)}{\partial \boldsymbol{\theta}_m}\right\vert_{\boldsymbol{\theta}_m = \hat{\boldsymbol{\theta}}_m(\mathbf{y})} \hat{\Sigma}_m  \left. \frac{\partial g_m(\boldsymbol{\theta}_m)}{\partial \boldsymbol{\theta}_m}\right\vert_{\boldsymbol{\theta}_m = \hat{\boldsymbol{\theta}}_m(\mathbf{y})}\right).
\label{eqn:normphi}
\end{equation}
The result is that the approximate model-averaged posterior distribution of $\boldsymbol{\phi}$ given by (\ref{eqn:star}) is a mixture of normal distributions where each component is given by (\ref{eqn:normphi}) and weighted by $\tilde{\pi}(m|\mathbf{y},\mathbf{d})$.

It may not always be possible to find a closed form for the density of the prior distribution of $\boldsymbol{\phi}$, evaluation of which is necessary for the calculation of the SI loss given by (\ref{eqn:siloss}). In these cases, we suggest approximating the prior distribution of $\boldsymbol{\theta}_m$ for each $m \in \mathcal{M}$ by a normal distribution (with mean $\boldsymbol{\mu}_m = \mathrm{E}(\boldsymbol{\theta}_m|m)$ and variance $\Psi_m = \mathrm{var}(\boldsymbol{\theta}_m|m)$). Now the prior density $\pi_{\phi}(\boldsymbol{\phi})$ can be approximated via
$$\tilde{\pi}_{\phi}(\boldsymbol{\phi}) = \sum_{m \in \mathcal{M}} \tilde{\pi}_{\phi}(\boldsymbol{\phi}|m) \pi(m),$$
where $\tilde{\pi}_{\phi}(\boldsymbol{\phi}|m)$ is the density of $\mathrm{N}\left(\boldsymbol{\mu}_m,\Psi_m\right)$. Similar to the posterior distribution, the model-averaged prior distribution of $\boldsymbol{\phi}$ is approximated by a mixture of normal distributions but where the weights are the true prior model probabilities.

It is well known \citep[e.g.][]{ChalonerVerdinelli} that in cases where there is no model uncertainty, $|\mathcal{M}|=1$, $g(\boldsymbol{\theta}) = \boldsymbol{\theta}$, and under a normal approximation to the posterior distribution the expected SI loss is equal to the objective function that defines pseudo-Bayesian D-optimality (a commonly used criterion in classical optimal design of experiments). In Appendix \ref{app:pb} we discuss the relationship between the normal-based approximation to the SI loss above and the pseudo-Bayesian D-optimal approximation. It is shown that the objective function for pseudo-Bayesian D-optimality is itself an approximation to the expectation of the normal-based approximation to the SI loss. This places the normal-based approximations as being a compromise between the computationally expensive DLMC approximation and the computationally cheap pseudo-Bayesian D-optimal approximation. Furthermore, in Section \ref{sec:log_reg}, we empirically compare the two approximations.   

\paragraph{Absolute error loss}
The absolute error (AE) loss \citep[e.g.][pages 79-80]{robert_2007} is given by
$$\lambda_{AE}(\boldsymbol{\phi},\mathbf{y},\mathbf{d}) = \sum_{j=1}^q |\phi_j - \mathrm{Q}(\phi_j|\mathbf{y},\mathbf{d})|,$$
where $Q(\phi_j|\mathbf{y},\mathbf{d})$ is the model-averaged posterior marginal median of $\phi_j$. If there is no model uncertainty, i.e. $|\mathcal{M}|=1$, then we can approximate the posterior median by $\tilde{\mathrm{Q}}(\phi_j|\mathbf{y},\mathbf{d}) = g_m(\hat{\boldsymbol{\theta}}_m(\mathbf{y}))_j$. However, if $|\mathcal{M}|>1$ then there is no closed form for the median (or any quantile) of a mixture of normal distributions. To overcome this problem we use simulation. We generate a sample $\left\{ \boldsymbol{\phi}^c \right\}_{c=1}^C$ from the approximate model-average posterior distribution of $\boldsymbol{\phi}$. We then approximate $\mathrm{Q}(\phi_j|\mathbf{y},\mathbf{d})$ by the corresponding sample median. The sample $\left\{ \boldsymbol{\phi}^c \right\}_{c=1}^C$ can be generated as follows
\begin{enumerate}
\item
Generate $\left\{m^c\right\}_{c=1}^C$ where model $m$ is chosen with probability $\tilde{\pi}(m|\mathbf{y},\mathbf{d})$.
\item
For $c=1,\dots,C$, complete the following steps
		\begin{enumerate}
		\item
		Generate $\boldsymbol{\theta}_{m^c}^c$ from $\mathrm{N}\left(\hat{\boldsymbol{\theta}}_m^c(\mathbf{y}),\hat{\Sigma}_{m^c}\right)$.
		\item
		Set $\boldsymbol{\phi}^c = g_{m^c}(\boldsymbol{\theta}_{m^c}^c)$.
		\end{enumerate}
\end{enumerate}

\paragraph{Squared error loss}
The squared error loss \citep[e.g.][pages 77-79]{robert_2007} is given by
$$\lambda_{SE}(\boldsymbol{\phi},\mathbf{y},\mathbf{d}) = \left(\boldsymbol{\phi} - \mathrm{E}_{\boldsymbol{\phi}|\mathbf{y},\mathbf{d}}\left[\boldsymbol{\phi}\right]\right)^T\left(\boldsymbol{\phi} - \mathrm{E}_{\boldsymbol{\phi}|\mathbf{y},\mathbf{d}}\left[\boldsymbol{\phi}\right]\right),$$
where
\begin{equation}
\mathrm{E}_{\boldsymbol{\phi}|\mathbf{y},\mathbf{d}}\left[\boldsymbol{\phi}\right] = \sum_{m \in \mathcal{M}} \mathrm{E}_{\boldsymbol{\theta}_m|\mathbf{y},m,\mathbf{d}}\left[g_m(\boldsymbol{\theta}_m)\right] \pi(m|\mathbf{y},\mathbf{d}).
\label{eqn:postmean}
\end{equation}
Typically, unless the $g_m$ are linear functions, the posterior mean $\mathrm{E}_{\boldsymbol{\theta}_m|\mathbf{y},m,\mathbf{d}}\left[g_m(\boldsymbol{\theta}_m)\right]$ will not be available in closed form. To approximate this posterior mean, we use the simulation approach described above for the case of approximating the posterior median, only replacing the sample median by sample mean. 

\subsubsection{Model discrimination} \label{sec:moddisc}

Now suppose the experimental aim is model discrimination and thus inference is based on the posterior model probabilities. Consider the following loss functions. In both cases, we can derive approximations by replacing the marginal likelihood, $\pi(\mathbf{y}|m,\mathbf{d})$, or posterior model probability, $\pi(m|\mathbf{y},\mathbf{d})$ by the corresponding approximations, $\tilde{\pi}(\mathbf{y}|m,\mathbf{d})$ or $\tilde{\pi}(m|\mathbf{y},\mathbf{d})$, given by (\ref{eqn:7a}) and (\ref{eqn:7b}), respectively.

\paragraph{0-1 loss}
The 0-1 loss \citep[e.g.][pages 80-81]{robert_2007} is given by
$$\lambda_{01}(m,\mathbf{y},\mathbf{d}) = 1 - I(m = M(m|\mathbf{y},\mathbf{d})),$$
where $I(A)$ denotes the indicator of event $A$ and $M(m|\mathbf{y},\mathbf{d})) = \mathrm{arg} \max_{m \in \mathcal{M}} \pi(\mathbf{y}|m,\mathbf{d})\pi(m)$ is the posterior modal model. The design that minimises the expected 0-1 loss equivalently maximises the expected posterior model probability of the modal model. 

\paragraph{Model self-information loss}
The model self-information loss \citep[e.g][]{mcgree_2015} is derived by extending the self-information loss for parameters to the posterior model probabilities. It is given by
$$\lambda_{MSI}(m,\mathbf{y},\mathbf{d}) = \log \pi(m) - \log \pi(m|\mathbf{y},\mathbf{d}).$$

%

\subsection{Logistic Regression} \label{sec:log_reg}
The setup for the following example is adapted from \cite{woods_overstall_2016} who correspondingly adapted it from a simpler problem studied by \cite{Woods2006} and \cite{gotwalt_jones_steinberg_2009}. It concerns a first-order logistic regression model in $k=4$ factors and $n$ runs. Although from a Bayesian inference perspective, logistic regression is a relatively simple model, it (or more generally some type of binary response model) is frequently used to benchmark new computational approaches in statistics \citep[e.g.][]{minka_2001,girolami_calderhead_2011,hoffman_gelman_2014}. Moreover, until \cite{woods_overstall_2016}, fully Bayesian design for such a model had not previously been attempted in the literature indicating that in the context of fully Bayesian design logistic regression remains a non-trivial problem.

A binary response is measured for $G$ blocks each of $n_G=6$ runs, i.e. the total number of runs is $n=Gn_G$. Let $y_{ij}$ and $x_{tij}$ denote the experimental response and value of the $t$th factor for the $j$th run from the $i$th block ($i=1,\dots,G$; $j=1,\dots,n_G$; $t=1,\dots,4$), respectively. It is assumed that $y_{ij} \sim \mathrm{Bernoulli}\left(\rho_{ij}\right)$, independently, where
\begin{eqnarray*}
\log \left( \frac{\rho_{ij}}{1-\rho_{ij}}\right) & = & \beta_0 + \gamma_{0i} + \sum_{t=1}^4 v_t \left(\beta_t + \gamma_{ti}\right) x_{tij},\\
& = & \mathbf{x}_{ij}^T \left(\mathbf{v} \circ \left(\boldsymbol{\beta} + \boldsymbol{\gamma}_i \right)\right),
\end{eqnarray*}
where $\boldsymbol{\beta} = (\beta_0,\beta_1,\beta_2,\beta_3,\beta_4)$ are the regression parameters, $\boldsymbol{\gamma} = \left(\boldsymbol{\gamma}_1,\dots,\boldsymbol{\gamma}_G\right)$ are the block-specific parameters, $\mathbf{v} = \left(v_0,\dots,v_4\right)$ is a binary vector with $v_t=1$ ($v_t=0$) indicating whether the $t$th factor is active (inactive), and $\circ$ denotes element-wise multiplication. The complete $p_m \times 1 $ vector of parameters is $\boldsymbol{\theta}_m = \left(\boldsymbol{\beta}_m,\boldsymbol{\gamma}_m\right)$. Each model $m \in \mathcal{M}$ is determined by the $|\mathcal{M}| = 2^4 = 16$ combinations of $\mathbf{v}$. 

Let $\mathbf{X} = \left(\mathbf{X}_1^T,\dots,\mathbf{X}_G^T\right)^T$ be the $n \times 5$ model matrix where $\mathbf{X}_i$ is the $n_G \times 5$ model matrix for the $i$th group with $j$th row given by $\mathbf{x}_{ij}^T$. The design is given by $\mathbf{d} = \mathrm{vec}(\mathbf{D}^T)$ where $\mathbf{D}$ is the $n \times 4$ matrix given by $\mathbf{X}$ with the first column of ones (corresponding to the intercept) removed. The design space $\mathcal{D}$ has dimensionality $W=4n$ and is such that each element of $\mathbf{d}$ lies in the interval $[-1,1]$.

From \cite{Woods2006} and \cite{woods_overstall_2016}, we assume independent prior distributions for each element of $\boldsymbol{\beta}$ with lower and upper limits given by $\mathbf{L} = (-3,4,5,-6,-2.5)$ and $\mathbf{U} = (3,10,11,0,3.5)$. Following \cite{woods_overstall_2016}, we assume two different prior distributions for each $\boldsymbol{\gamma}_i$ ($i=1,\dots,G$).
\begin{enumerate}
\item[(i)]
A prior point mass at $\boldsymbol{\gamma}_i = \mathbf{0}$ for all $i$, resulting in standard logistic regression with $p_m = 1 + \sum_{t=1}^4 v_t$.
\item[(ii)]
A hierarchical prior distribution in which elements of $\boldsymbol{\gamma}_i$ are independent and identically distributed as $\gamma_{ti} \sim \mathrm{U}[-\zeta_t,\zeta_t]$, for $t=0,\dots,4$ with $\zeta_t \in (0,Z_t)$ unknown and having triangular prior density $\pi(\zeta_t) = 2(Z_t-\zeta_t)/Z_t^2$ with $\left(Z_0,Z_1,\dots,Z_4\right) = \left(3,3,3,1,1\right)$. The numbers of parameters is $p_m = 1 + G +(1+G)\sum_{t=1}^4 v_t$.
\end{enumerate}

Under each of these two prior distributions, we find designs for the experimental aims of parameter estimation and model discrimination.

\subsubsection{Parameter Estimation} \label{sec:logregest}

We set $v_t = 1$ for all $t$ so that there is no model uncertainty and consider generating designs under the aim of estimating $\boldsymbol{\phi} = \boldsymbol{\beta}$. This means $g_m = g$ is a linear function given by $g(\boldsymbol{\theta}) = A \boldsymbol{\theta}$, where for i) standard logistic regression, $A = I_5$; and for ii) hierarchical logistic regression, $A = \left(I_5,R\right)$ is a $5 \times 5(1+G)$ matrix where $R$ is a $5 \times 5G$ matrix of zeros. 

We consider the self-information and squared error loss functions and compare designs found under the DLMC approximation (as found by \citealt{woods_overstall_2016} and referred henceforth as DLMC designs) against designs found under the NBMC approximations proposed in this paper (henceforth referred to as NBMC designs). To make comparisons valid, we found the NBMC designs under exactly the same implementation of ACE as those used to find the DLMC designs (see Appendix~\ref{app:ace} for details). Additionally, we also compare against pseudo-Bayesian D- and A-optimal designs (also as found by \citealt{woods_overstall_2016}).

We compare designs using relative efficiency. Let $\mathbf{d}_{SI}^*$ and $\mathbf{d}_{SE}^*$ be the DLMC designs under the SI and SE loss functions, respectively, found by \cite{woods_overstall_2016}. The relative SI and SE efficiencies of a design $\mathbf{d}$ are defined as
\begin{eqnarray}
R_{SI}(\mathbf{d}) & = & \frac{L_{SI}(\mathbf{d})}{L_{SI}(\mathbf{d}^*_{SI})} \times 100\%, \label{eqn:R1}\\
R_{SE}(\mathbf{d}) & = & \frac{L_{SE}(\mathbf{d}^*_{SI})}{L_{SI}(\mathbf{d})} \times 100\%, \label{eqn:R2}
\end{eqnarray}
respectively, where $L_{SI}$ and $L_{SE}$ refer to the expected SI and SE loss functions, respectively. Note that the definition of relative SI efficiency, given by (\ref{eqn:R1}), follows from how the expected SI loss is non-positive. The relative efficiencies are approximated by approximating the expected losses in (\ref{eqn:R1}) and (\ref{eqn:R2}) by using the DLMC approximation. 

\begin{figure}
\centering
\includegraphics[scale=0.7]{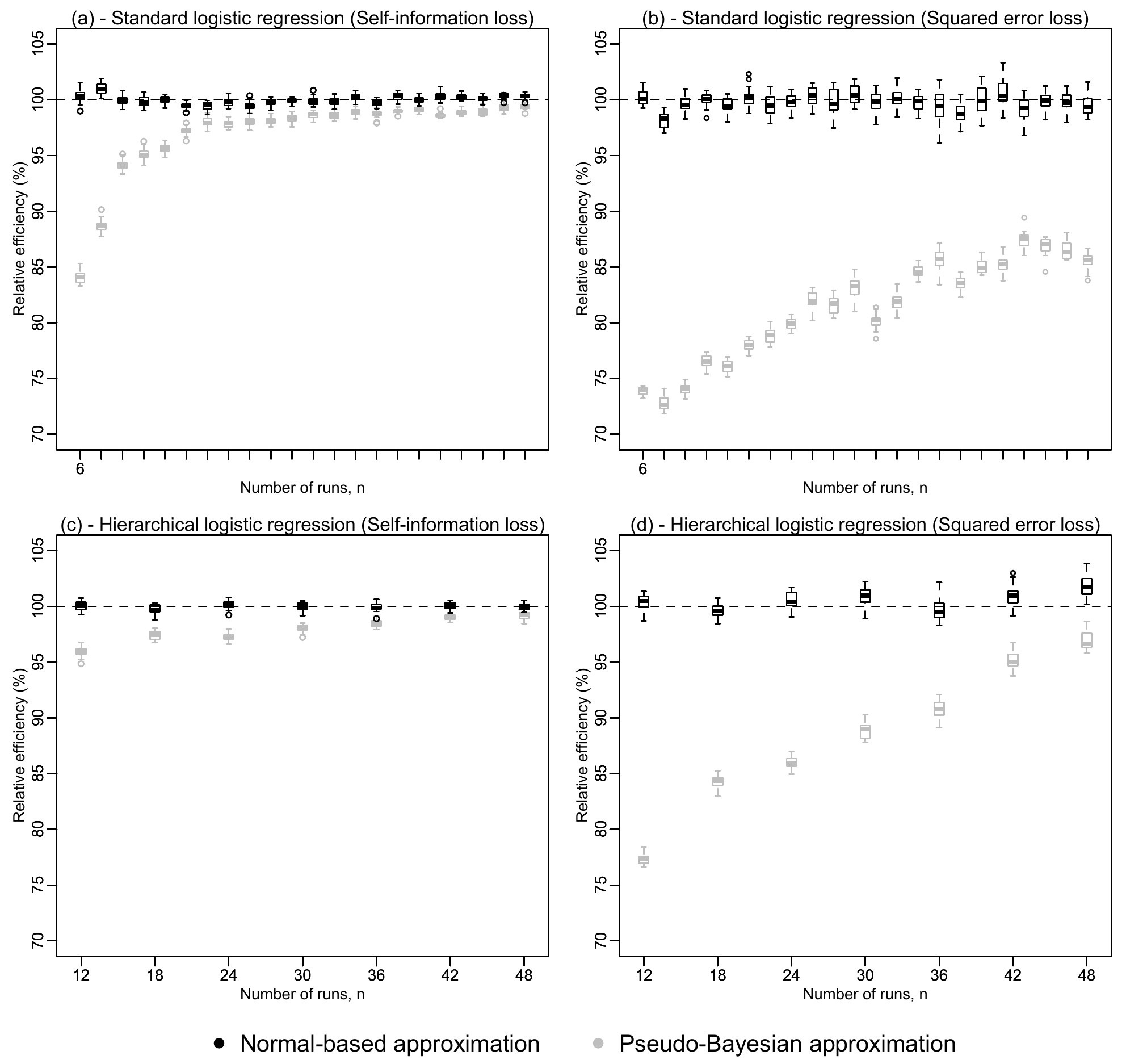}
\caption{\label{fig:log_reg_est} Boxplots of twenty DLMC approximations ($B=20,000$) to the relative self-information ((a) and (c)) and squared error ((b) and (d)) efficiencies plotted against $n \in N_S= \left\{6,8,\dots,48\right\}$ for standard logistic regression ((a) and (b)) and against $n \in N_H = \left\{12,18, \dots,48\right\}$ for hierarchical logistic regression ((c) and (d)).}
\end{figure}

The top row of Figure~\ref{fig:log_reg_est} shows boxplots of twenty DLMC approximations to the relative efficiency for SI (left) and SE (right) for the NBMC and pseudo-Bayesian designs plotted against $n \in N_S = \left\{6,8,\dots,48\right\}$ (meaning $W$ ranges from 24 to 192) for standard logistic regression. The bottom row shows the same boxplots for hierarchical logistic regression plotted against $n \in N_H = \left\{12,18, \dots,48\right\}$ (meaning $W$ ranges from 48 to 192). In both cases, the relative efficiencies of the NBMC designs are clearly very close to one for all values of $n$ indicating that the performance of these designs (in terms of expected loss) is very close to the performance of the DMLC designs. However, the relative efficiency of the pseudo-Bayesian designs only become close to one as $n$ gets larger. In the case of the SI loss, it appears from Figure~\ref{fig:log_reg_est} that we could obtain a negligible difference in expected SI loss for values of $n$ over approximately forty by using the pseudo-Bayesian D-optimal design. This design is computationally more efficient to find than a fully Bayesian design, however, knowing that it had nearly equivalent performance to the fully Bayesian design would be hard without first finding the fully Bayesian design.

\begin{figure}
\centering
\includegraphics[scale=0.7]{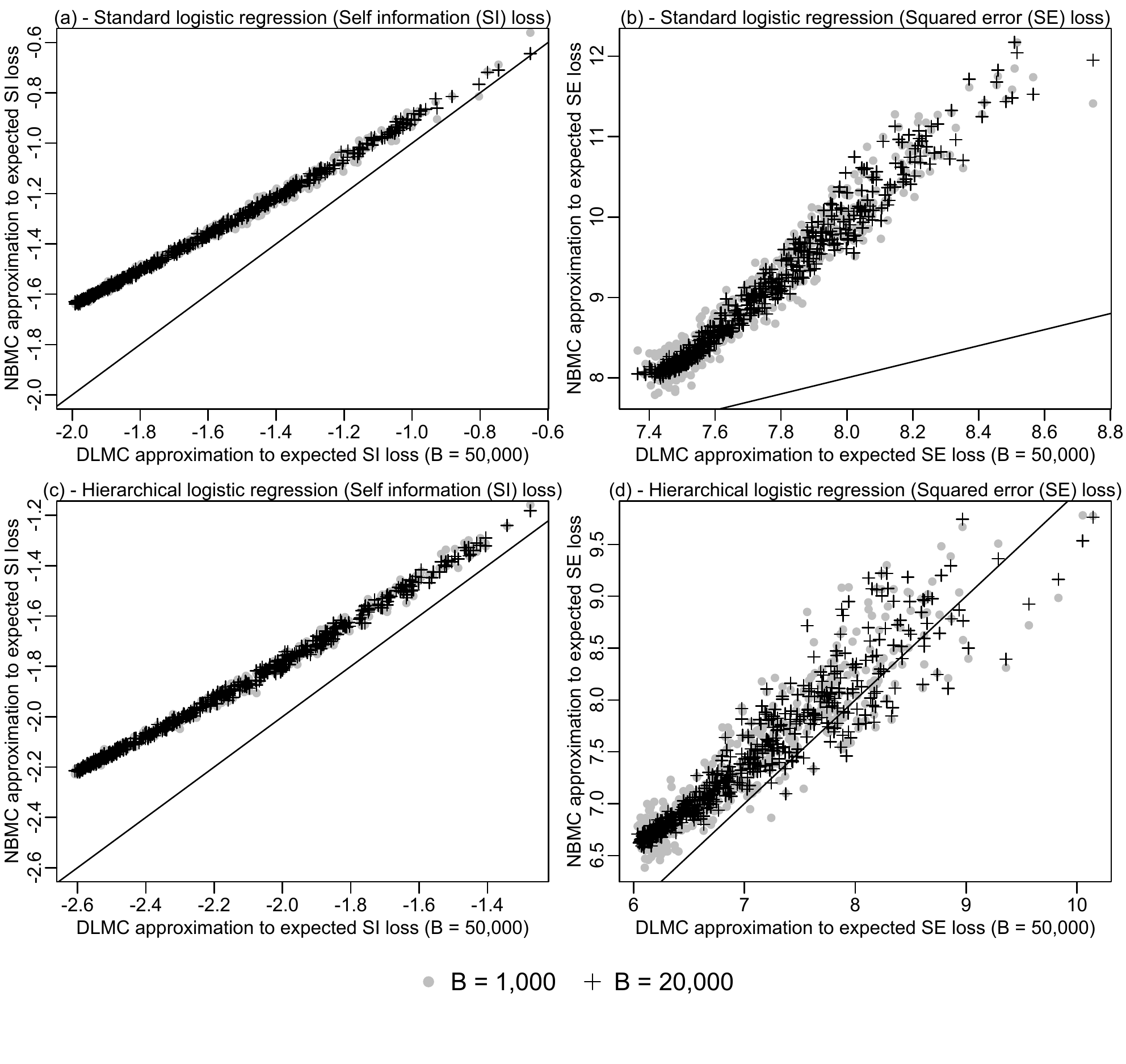}
\caption{\label{fig:log_reg_comp}  Plots of NBMC approximations (with B=1,000 and $B=20,000$) to the expected loss plotted against the DLMC approximation ($B=50,000$) to the expected loss for standard ((a) and (b) for $n=6$) and hierarchical ((c) and (d) for $n=12$) logistic regression under the self-information ((a) and (c)) and squared error ((b) and (d)) loss function. A line through the origin with slope one has been added to aid in the comparison.}
\end{figure}

We now investigate the accuracy of the NBMC approximation to the expected loss. For the SI and SE loss, we randomly generate $T$ designs and for each one calculate three different approximations to the expected loss: a DLMC approximation with $B=50000$ (which we consider near exact evaluation of the expected loss), and NBMC approximations with $B=1000$ and $B=20000$. The $t$th design for $t=1,\dots,T$, is generated by perturbing the DLMC designs under each of the loss functions as
follows
\begin{eqnarray*}
\mathbf{d}_{SI}^{(t)} & = & (1-u_t)\mathbf{d}^*_{SI} + u_t \mathbf{d}^{(t)},\\
\mathbf{d}_{SE}^{(t)} & = & (1-u_t)\mathbf{d}^*_{SE} + u_t \mathbf{d}^{(t)}
\end{eqnarray*}
where, at each iteration, $u_t \sim \mathrm{U}(0,\frac{1}{2})$, and $\mathbf{d}^{(t)}$ is a design in which each element is generated from $\mathrm{U}(-1,1)$.  The top row of Figure~\ref{fig:log_reg_comp} shows plots of the two NBMC approximations to the expected loss plotted against the DLMC approximation to the expected loss for standard logistic regression under the SI (left) and SE (right) loss function for $n=6$ (i.e. the smallest number of runs considered). The bottom row of Figure~\ref{fig:log_reg_comp} shows the same plots for hierarchical logistic regression with $n=12$ (again the smallest number of runs considered). In all four cases, although the NBMC approximation to the expected loss can be inaccurate, especially for designs close to the minimum, the ordering of designs in terms of expected loss is the same as for the DLMC approximation. This means the NBMC approximation to the expected loss is minimised for design close to the design that minimises the DLMC approximation.

\begin{figure}
\centering
\includegraphics[scale=0.7]{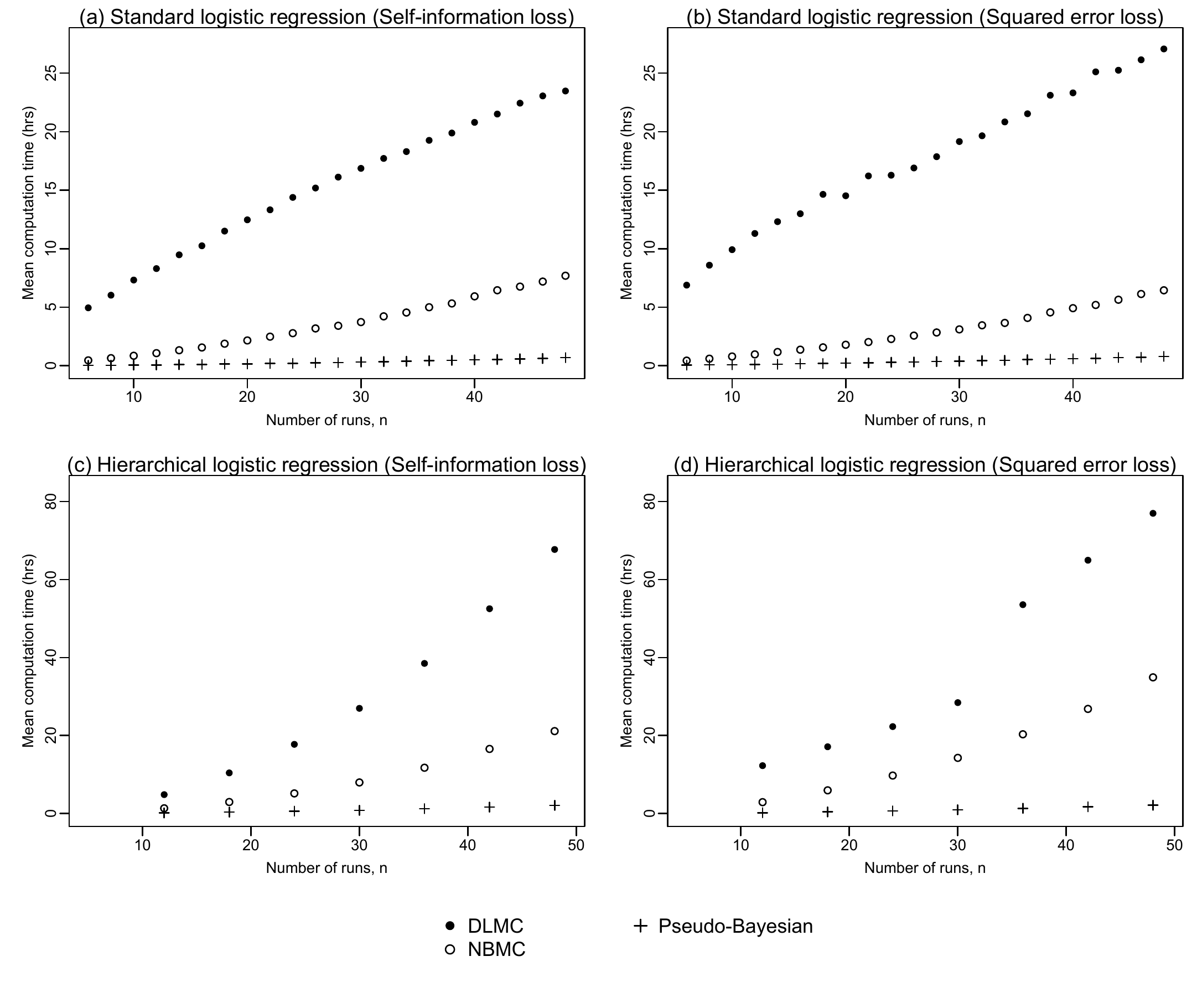}
\caption{\label{fig:log_reg_timings}  Plots of mean computer time required to find the NBMC, DLMC and pseudo-Bayesian designs for standard ((a) and (b)) and hierarchical ((c) and (d)) logistic regression under the SI ((a) and (c)) and SE ((b) and (d)) loss function.}
\end{figure}

Figure~\ref{fig:log_reg_timings} shows plots of the mean computer time required to compute the three types of design (NBMC, DLMC and pseudo-Bayesian) for the models and loss functions considered. Note that the algorithm was run on the IRIDIS 4 supercomputer facility at the University of Southampton which has 2.6Ghz processors with 4Gb memory. Note that the MBMC designs are typically found in a third of the computing time required to find the DLMC designs. The pseudo-Bayesian designs require the smallest amount of computing time but this should be judged in parallel with the lack of efficiency these designs exhibit when compared to the corresponding fully Bayesian design (see Figure~\ref{fig:log_reg_est}).

\subsubsection{Model discrimination} \label{sec:logregsel}

\begin{figure}
\centering
\includegraphics[scale=0.8, viewport = 0pt 230pt 500pt 510pt, clip=true]{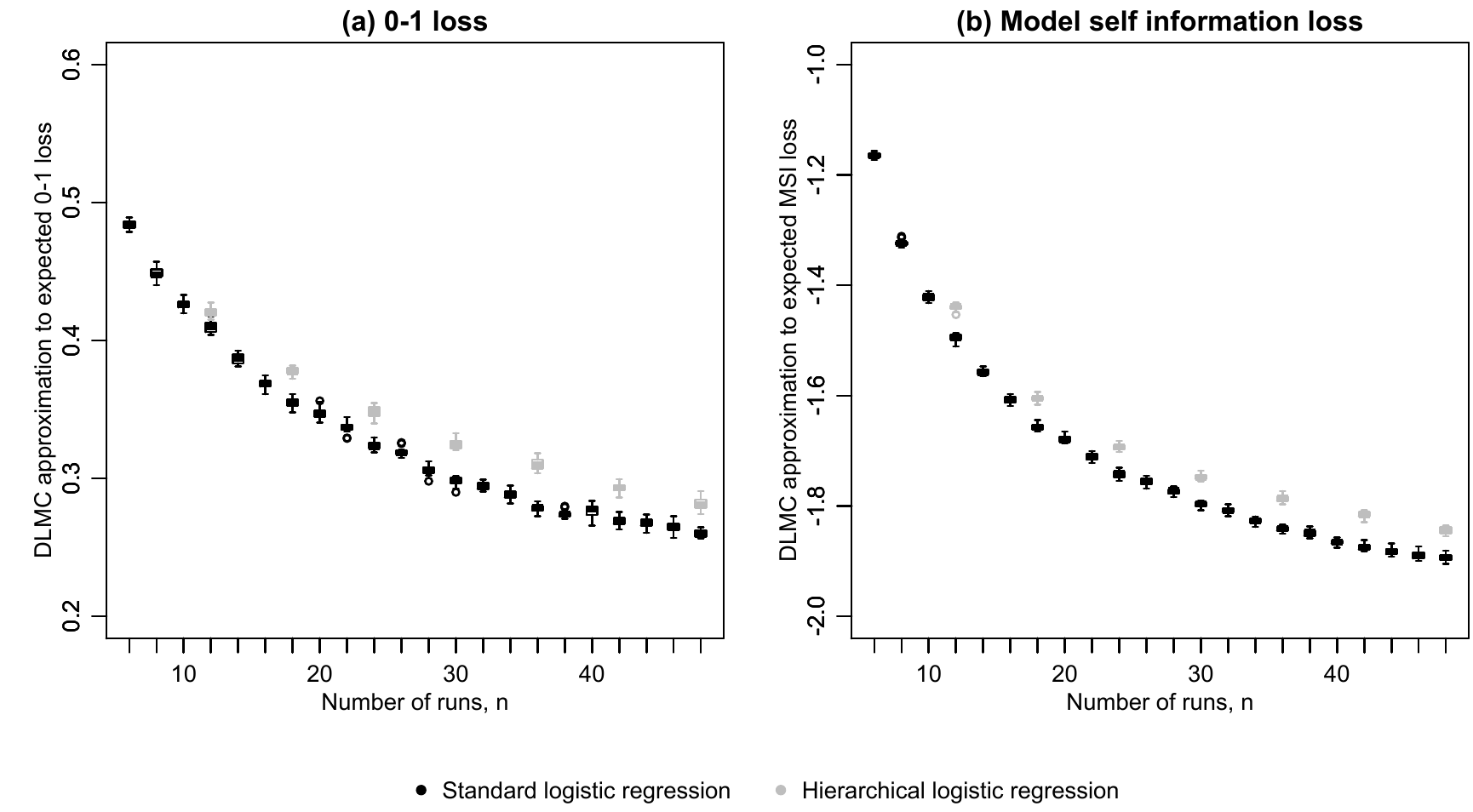}
\caption{\label{fig:log_reg_sel} Plots of the DLMC approximation ($B=20,000$) to the expected 0-1 (a) and model self-information (b) loss function for standard and hierarchical logistic regression against $n$.}
\end{figure}

Now consider the aim of model discrimination. Suppose that $v_0=1$ (corresponding to the intercept) but that $v_t$ is unknown for $t=1,\dots,4$. Each unique combination of $(v_1,\dots,v_4)$ corresponds to a unique model $m$. The total number of models is $|\mathcal{M}| = 2^4 = 16$ ranging from $(v_1,\dots,v_4) = (0,0,0,0)$ (intercept only; all factors inactive) to $(v_1,\dots,v_4) = (1,1,1,1)$ (full model; all factors active). The prior model probabilities are such that $\mathrm{P}(v_t = 1)\sim\mathrm{U}[0,1]$ independently, for each $t=1,\dots,4$. For model $m$, let $(v_{m1},\dots,v_{m4})$ denote the values of $v_t$ where the number of regression parameters is given by $b_m = 1 + \sum_{t=1}^4 v_{mt}$. The marginal prior model probabilities are then given by 
$$\pi(m) = \frac{1}{5 \genfrac{(}{)}{0pt}{}{4}{b_m-1}},$$
which corrects for Bayesian multiplicity \citep{scott_berger_2010}.

We find designs for both standard and hierarchical logistic regression for each value of $n$ ($N_S$ for standard, and $N_H$ for hierarchical), under both the 0-1 and model self-information loss functions, respectively. Bayesian optimal design on such a scale for the experimental aim of model discrimination has not been addressed previously in the literature. It would be infeasible to find DLMC designs in this situation since it would require $|\mathcal{M}| = 16$ Monte Carlo approximations to the marginal likelihood for every $b=1,\dots,B$. Conversely, the normal-based approximations will require $|\mathcal{M}| = 16$ Laplace approximations which will be computationally less intensive.

However, although it is infeasible to use the DLMC approximation to find designs, we can use it to assess the NBMC designs. Figure~\ref{fig:log_reg_sel} shows boxplots of twenty DLMC approximations to the expected 0-1 and model self-information loss functions for both standard and hierarchical logistic regression against $n$ for the NBMC designs. Note that the expected loss for the hierarchical model is always greater than for the standard model and that this difference increases as $n$ increases. This is due to the extra uncertainty introduced by the blocks and their associated block-specific parameters (whose number is proportional to  $n$).

\begin{figure}
\centering
\includegraphics[scale=0.8]{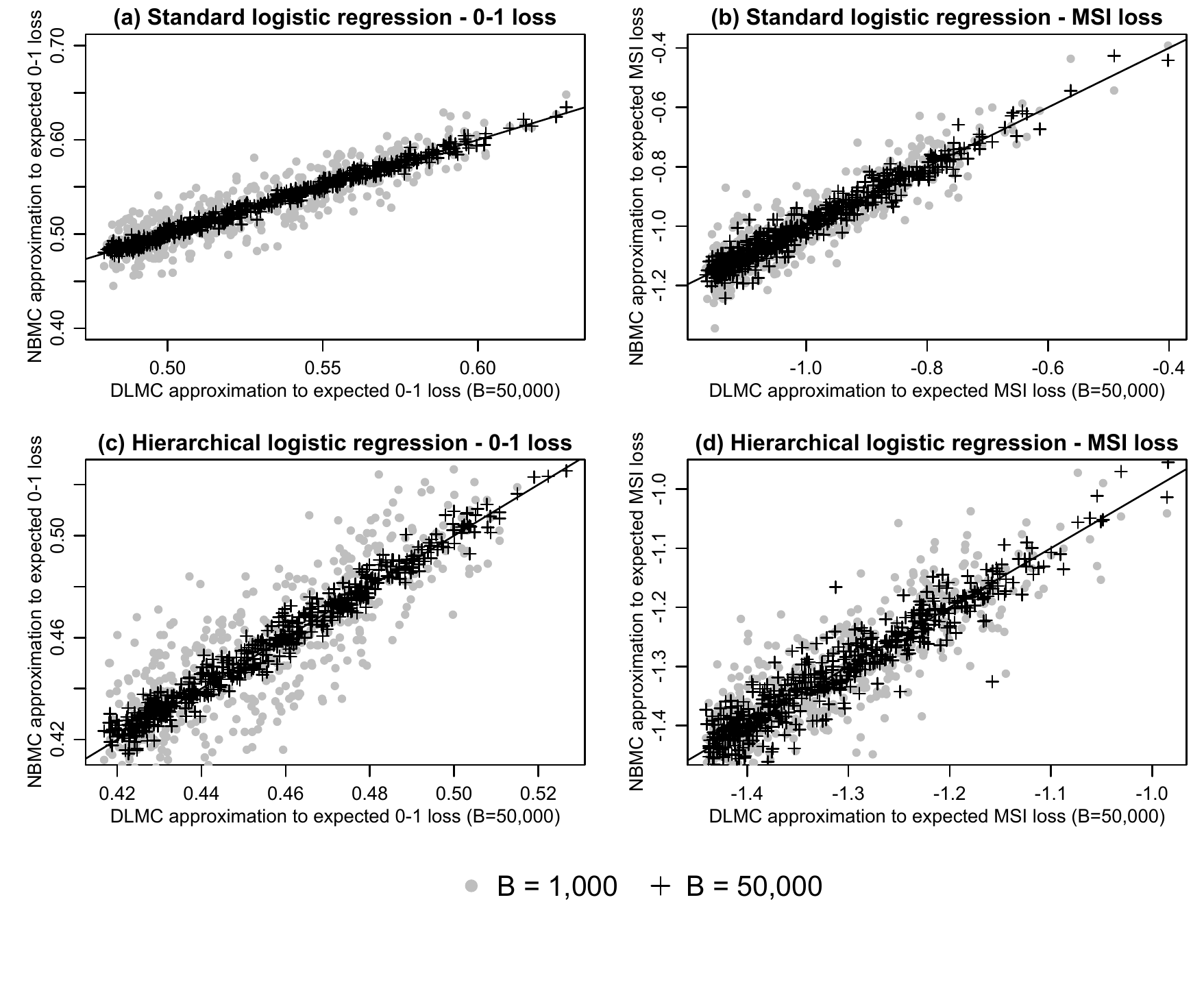}
\caption{\label{fig:logregselcomp} Plots of NBMC approximation (with $B=1,000$ and $B=20,000$) to the expected 0-1 ((a) and (c)) and model self-information ((b) and (d)) loss plotted against the corresponding DLMC approximation ($B=50,000$) for standard ((a) and (b) for $n=6$) and hierarchical ((c) and (d) for $n=12$) logistic regression.}
\end{figure}

Similar to Section~\ref{sec:logregest} we check the validity of the approximation by plotting NBMC approximations (with $B=1,000$ and $B=20,000$) against the DLMC approximation to the expected loss with $B=50,000$. Figure~\ref{fig:logregselcomp} shows the resulting plots for $n=6$ (standard logistic regression) and $n=12$ (hierarchical logistic regression). In this case, we can see that the NBMC approximations to the expected loss appear very accurate for both loss functions.

\subsection{Mechanistic Modelling of Chemical Reactions} \label{sec:BH}

In this section we consider the famous example from \cite{box_hill_1967} concerning discriminating between non-linear models for describing chemical reactions. Suppose the $i$th run consists of specifying temperature $x_{i1} \in (0,150)$ and reaction time $x_{i2} \in (450,600)$, i.e. $\mathbf{d}_i = (x_{i1},x_{i2})$, and measuring the yield $y_i$ from a chemical reaction, for $i=1,\dots,n$. It is assumed that
$$y_i \sim \mathrm{N}\left(\eta_m(\boldsymbol{\theta};\mathbf{d}_i), \sigma^2 \right),$$
independently for $i=1,\dots,n$ where $m \in \mathcal{M}$. Consider the set $\mathcal{M} = \left\{1,2,3,4\right\}$ of $|\mathcal{M}| = 4$ competing models for $\eta_m(\boldsymbol{\theta};\mathbf{d}_i)$ given as follows
$$\begin{array}{lclcllclcl}
m=1& : & \eta_1(\boldsymbol{\theta};\mathbf{d}_i) & = & \exp \left( - \theta_{11}x_{i1} \exp\left( - \frac{\theta_{12}}{x_{i2}}\right)\right); \qquad & m=2 & : & \eta_2(\boldsymbol{\theta};\mathbf{d}_i) &= & 1/ \left( 1+ \theta_{21}x_{i1} \exp\left( - \frac{\theta_{22}}{x_{i2}}\right)\right);\\
m=3& : &\eta_3(\boldsymbol{\theta};\mathbf{d}_i)& = &1/ \left( 1+ \theta_{31}x_{i1} \exp\left( - \frac{\theta_{32}}{x_{i2}}\right)\right)^{\frac{1}{2}}; \qquad & m=4 & : & \eta_4(\boldsymbol{\theta};\mathbf{d}_i) & = & 1/ \left( 1+ \theta_{41}x_{i1} \exp\left( - \frac{\theta_{42}}{x_{i2}}\right)\right)^{\frac{1}{3}};
\end{array}$$
where the unknown parameters $\boldsymbol{\theta}_m = (\theta_{m1},\theta_{m2})$ have the same interpretation under each model. Following \cite{box_hill_1967}, a-priori we assume that $\theta_{m1} \sim \mathrm{N}(400,25^2)$ and $\theta_{m2} \sim \mathrm{N}(5000,250^2)$, independently, and that $\pi(m) = 1/4$. \cite{box_hill_1967} assumed that the response variance was fixed as $\sigma^2 =0.1^2$. However, we let $\sigma^2$ be unknown and assume that $\sigma^2 \sim \mathrm{U}[0,1]$. 

We consider $n = \left\{5,10,15,\dots,50\right\}$ meaning that $W$ ranges from 10 to 100. To demonstrate the versatility of the normal-based approximations presented in this paper, for each value of $n$, we find NBMC designs under a total of eight different loss functions. We consider the three exemplar loss functions (Section~\ref{sec:paraest}) for parameter estimation for a) $\boldsymbol{\phi} = g_m(\boldsymbol{\theta}_m) = \boldsymbol{\theta}_m$; and b) $\boldsymbol{\phi} = g_m(\boldsymbol{\theta}_m) = \theta_{m2}/\theta_{m1}$, where in the latter case we are interested in the ratio of the two unknown parameters. In both cases we are taking account of model uncertainty by considering the model-averaged posterior distribution of $\boldsymbol{\phi}$. We also consider the two exemplar loss functions for model discrimination (Section~\ref{sec:moddisc}). 

\begin{figure}
\centering
\includegraphics[scale=0.75,]{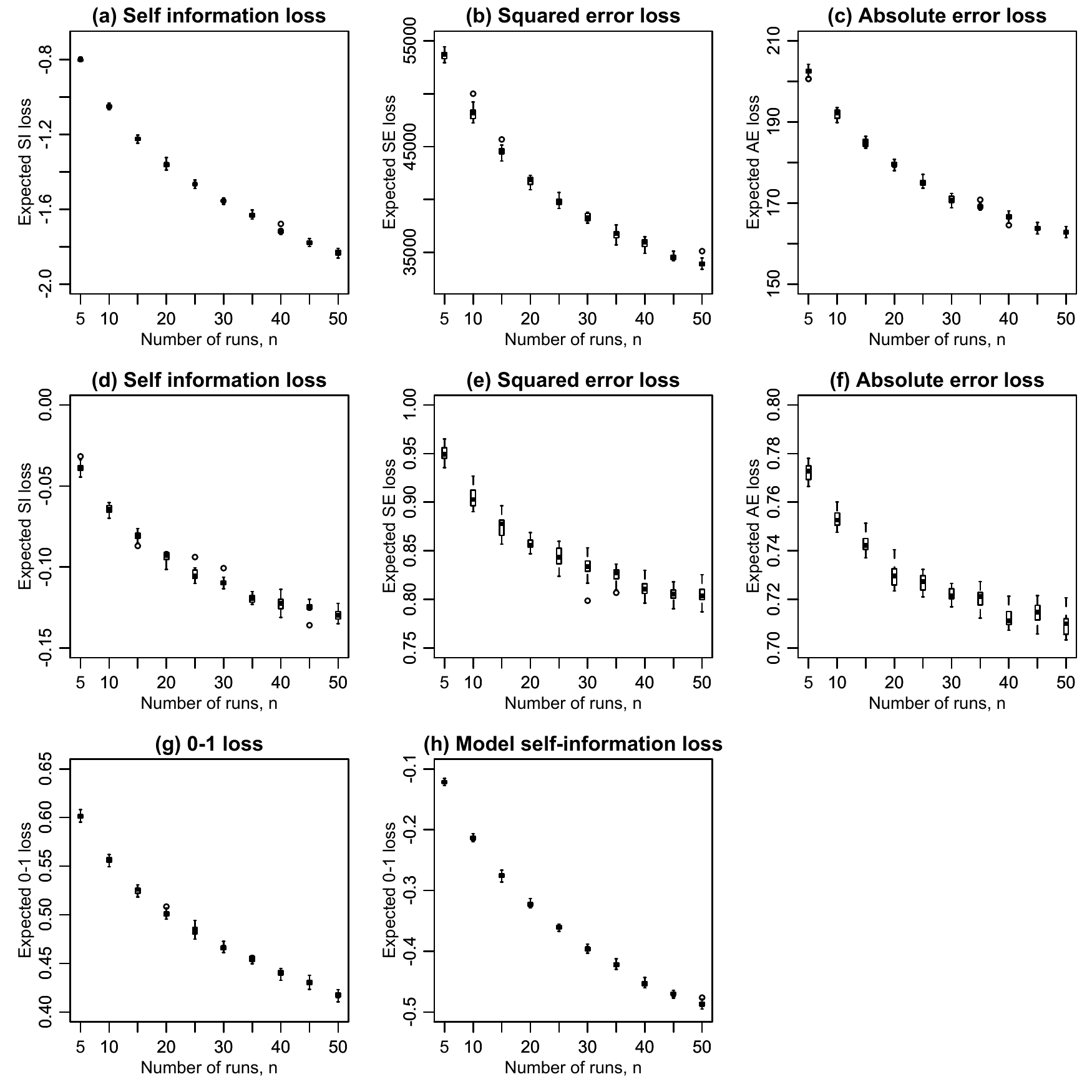}
\caption{\label{fig:bh} Boxplots of twenty NBMC approximations ($B=20,000$) to the expected loss for the NBMC designs plotted against $n$ under the each of the loss functions for the non-linear model example.}
\end{figure}

It is not clear how to find DLMC approximations to the expected loss for every loss function considered in this section. For example, a DLMC approximation to the model-averaged posterior median required for the AE loss would require samples from the posterior distribution of $\boldsymbol{\theta}_m$ for each $m \in \mathcal{M}$ and $\mathbf{y}^b$, for $b=1,\dots,B$. Therefore in this section we rely on the NBMC approximations to assess the performance of designs. Figure~\ref{fig:bh} shows twenty boxplots of the NBMC approximation ($B=20,000$) to the expected loss for the NBMC designs plotted against $n$ for each of the loss functions. Note how the expected loss functions for $\boldsymbol{\phi} = \boldsymbol{\theta}_m$ have a faster relative decrease in expected loss with increasing $n$ than the expected loss for $\phi = \theta_{m2}/\theta_{m1}$. This is easiest to see for the SE in loss in Figure~\ref{fig:bh}(b) and~\ref{fig:bh}(e), where the expected SE loss for $\boldsymbol{\phi} = \boldsymbol{\theta}_m$ has a relative decrease of approximately 35\% when $n$ increase from $5$ to $50$. The corresponding relative decrease in expected SE loss for $\phi = \theta_{m2}/\theta_{m1}$ is approximately 15\%. This is due to the form of the parameterisation of interest $g_m(\boldsymbol{\theta}_m)$. When $\boldsymbol{\phi} = \boldsymbol{\theta}_m$, the trace of the prior variance of $\phi$ is 63,125 which gives an upper bound on the expected loss. The corresponding value for $\phi = \theta_{m2}/\theta_{m1}$ is approximately 1. Under the latter parameterisation, it appears the choice of design does not have as great an impact on the expected SE loss than the former.

\section{Conclusion}\label{sec:con}

In this paper, we have proposed the use of normal-based approximations to posterior summaries to aid in the approximation of the loss function. The resulting approximate loss can be used in conjunction with any algorithm suitable for finding the design that minimises the expected loss function. The methodology was used in conjunction with the ACE algorithm to find designs which have similar performance (in terms of expected loss) to designs generated by the original ACE algorithm with the DLMC approximation to the expected loss, but using a fraction of the computing time. The methodology was also able to find designs for problems, under model uncertainty, where use of the DLMC approach would be infeasible and, as such, have not previously been addressed in the literature.

The normal-based approximations utilised in this paper are formed by using the location of, and curvature around, the posterior mode. Taking advantage of alternative normal-based of other deterministic approximations may be of future interest.  In particular, one could consider using the expectation propagation (EP) algorithm of \cite{minka_2001} to form efficient approximations to the posterior distribution of the parameters.  Similar to mode/curvature based approximations used in this paper, EP also provides an efficient approximate of the marginal likelihood.  EP could potentially be useful in a wider range of problems as any distribution in the exponential family could be considered as the parametric form for approximating the posterior distribution.

\section{Acknowledgments}
The authors would like to thank the two anonymous referees for their insightful comments which have significantly improved the paper. The authors would also like to thank Prof David Woods (University of Southampton) for helpful suggestions on a draft of this paper. This work arose from discussions and collaborations initiated at the ``Bayesian Optimal Design of Experiments'' workshop (Queensland University of Technology, Brisbane, Australia, December 2015; \texttt{http://www.bode2015.wordpress.com}). A.M. Overstall was supported by a Research Incentive Grant (70424) from the Carnegie Trust for the Universities of Scotland. C.C. Drovandi was supported by an Australian Research Council's Discovery Early Career Researcher Award funding scheme (DE160100741).

\appendix

\section{Details on the approximate coordinate exchange algorithm} \label{app:ace}

\subsection{ACE algorithm}

\begin{enumerate}
\item
Choose an initial design $\mathbf{d}^0 = \left(d_1^0,\dots,d_W^0\right)$ and set the current design to be $\mathbf{d}^C = \left(d_1^C,\dots,d_W^C\right)=\mathbf{d}^0$. 
\item
For $i=1,\dots,W$ complete the following steps
\begin{enumerate}
\item
Let $L^i(d) = L(d^C_1,\dots,d^C_{i-1},d,d^C_{i+1},\dots,d^C_W)$ be the function given by the expected loss function which only varies over the design space, $\mathcal{D}_i$, for the $i$th element.
\item \label{alg:approx} 
For $j=1,\dots,Q$, evaluate the MC approximation to the expected loss given by
$$z_j = \hat{L}^i(d_j),$$
for $\left\{d_1,\dots,d_Q\right\} \in \mathcal{D}_i$. Fit a GP emulator to $\left\{z_j,d_j\right\}_{j=1}^Q$ and set $\tilde{L}^i(d)$ to be the resulting predictive mean.
\item
Find
$$d_i^* = \mathrm{arg} \mathrm{min}_{d \in \mathcal{D}_i} \tilde{L}^i(d),$$
and let $\mathbf{d}^* = \left(d^C_1,\dots,d^C_{i-1},d^*,d^C_{i+1},\dots\dots,d^C_W\right)$ be the proposed design.
\item \label{donaldtrump}
Set $\mathbf{d}^C = \mathbf{d}^*$ with probability $p^*$.
\end{enumerate}
\item
Return to step 2.
\end{enumerate}

\subsection{Comparison procedure}

In step \ref{donaldtrump}, we accept the proposed design, $\mathbf{d}^*$ with probability $p^*$. The proposed design originates from from the Gaussian process emulator. Similar to all statistical models, Gaussian process emulators, can fit inadequately. To mitigate the effects of an inadequate emulator, \cite{woods_overstall_2016} proposed a comparison between the proposed design $\mathbf{d}^*$ and the current design $\mathbf{d}^C$. Note that the proposed design $\mathbf{d}^*$ should be accepted if 
\begin{equation}
\mathrm{E}_{\boldsymbol{\theta}_m,m,\mathbf{y}|\mathbf{d}^*}\left[ \lambda (\boldsymbol{\theta}_m,m,\mathbf{y},\mathbf{d}^*) \right] < \mathrm{E}_{\boldsymbol{\theta}_m,m,\mathbf{y}|\mathbf{d}^C}\left[ \lambda (\boldsymbol{\theta}_m,m,\mathbf{y},\mathbf{d}^C) \right]
\label{eqn:compwert}
\end{equation}
For $b=1,\dots,B$ we generate samples $\left\{\lambda^b_*\right\}_{b=1}^B$ and $\left\{\lambda^b_C\right\}_{b=1}^B$ as follows
\begin{eqnarray*}
\lambda^b_* & = & \lambda (\boldsymbol{\theta}_m^{*b},m^{*b},\mathbf{y}^{*b},\mathbf{d}^*),\\
\lambda^b_C & = & \lambda (\boldsymbol{\theta}_m^b,m^b,\mathbf{y}^b,\mathbf{d}^C),
\end{eqnarray*}
where $\left\{\boldsymbol{\theta}_m^{*b},m^{*b},\mathbf{y}^{*b}\right\}_{b=1}^B$ and $\left\{\boldsymbol{\theta}_m^{b},m^{b},\mathbf{y}^{b}\right\}_{b=1}^B$ are samples from the joint distribution of $\boldsymbol{\theta}_m$, $m$ and $\mathbf{y}$ conditional on $\mathbf{d}^*$ and $\mathbf{d}^C$, respectively. We use these samples to perform a Bayesian hypothesis test of (\ref{eqn:compwert}). The form of the Bayesian hypothesis test, as described in \cite{woods_overstall_2016}, assumes that the $\lambda^b_*$'s and $\lambda^b_C$'s are continuous and their distribution reasonably assumed normal. In this case, the probability of accepting the proposed design is
$$p^* = 1 - F\left(-\frac{\sum_{b=1}^B \lambda_C^b - \sum_{b=1}^B \lambda_*^b}{\sqrt{2B \hat{v}}}\right),$$
where $F(\cdot)$ is the distribution function of the $t$-distribution with $2B-2$ degrees of freedom,
$$\hat{v} = \frac{\sum_{b=1}^B (\lambda_C^b - \bar{\lambda}_C)^2 + \sum_{b=1}^B (\lambda_*^b - \bar{\lambda}_*)^2}{2B - 2},$$
and $\bar{\lambda}_C$ and $\bar{\lambda}_*$ are the sample means of the $\lambda_C^b$'s and $\lambda_*^b$'s, respectively.

The assumption of normality will clearly be violated for the 0-1 loss function for model discrimination, described in Section~\ref{sec:losses}, where the $\lambda^b_*$'s and $\lambda^b_C$'s will be binary in the set $\left\{0,1\right\}$. For such loss functions, we introduce the following modification. Assume that 
\begin{eqnarray*}
\lambda_C^b & \stackrel{\mathrm{iid}}{\sim} & \mathrm{Bernoulli}\left(\rho_C\right),\\
\lambda_*^b & \stackrel{\mathrm{iid}}{\sim} & \mathrm{Bernoulli}\left(\rho_*\right),
\end{eqnarray*}
for $b=1,\dots,B$. We also assume the following independent prior distributions: $\rho_C \sim \mathrm{U}[0,1]$ and $\rho_* \sim \mathrm{U}[0,1]$. The resulting posterior distributions are
\begin{eqnarray*}
\rho_C|\lambda_C^1,\dots,\lambda_C^B & \sim & \mathrm{Beta}\left(1+B\bar{\lambda}_C,1+B-B\bar{\lambda}_C\right),\\
\rho_*|\lambda_*^1,\dots,\lambda_*^B & \sim & \mathrm{Beta}\left(1+B\bar{\lambda}_*,1+B-B\bar{\lambda}_*\right).
\end{eqnarray*}
The probability of accepting the new design is then given by
$$p^*  = \mathrm{P}\left(\rho_* < \rho_C |\lambda_C^1,\dots,\lambda_C^B,\lambda_*^1,\dots,\lambda_*^B\right)$$
which is evaluated via simulation as follows
$$p^* \approx \frac{1}{B} \sum_{b=1}^B F\left(\rho_C^b;1+B\bar{\lambda}_*,1+B-B\bar{\lambda}_*\right),$$
where $F\left(\cdot;a,b\right)$ denotes the distribution function of the $\mathrm{Beta}(a,b)$ and $\left\{\rho_C^b \right\}_{b=1}^B$ is a sample from \newline
$\mathrm{Beta}\left(1+B\bar{\lambda}_C,1+B-B\bar{\lambda}_C\right)$.


\subsection{Implementation details}

To reduce the likelihood of the ACE algorithm converging to local optima, it is restarted from $E$ different starting designs. These $E$ repetitions of the ACE algorithm aree run in an embarrassingly parallel fashion. Note that there is further scope for parallelising the ACE algorithm. The $Q$ evaluations of the Monte Carlo approximation to the expected loss in step \ref{alg:approx} could be parallelised. Furthermore, the calculation of the posterior mode for $b=1,\dots,B$ could also be parallelised. These have not been pursued here through. Even by repeating the ACE algorithm $E$ times does not guarantee that it will converge to the true optimal design. Following \cite{woods_overstall_2016} we set $E=20$, $Q=20$, and $B=1,000$, except for in the comparison procedure where $B=20,000$. These values were found by \cite{woods_overstall_2016} to perform well for a variety of examples. Additionally, \cite{MullerParmigiani} also found that $B$ could be similarly small when using their smoothing-based approach. Note that in their use of DLMC approximation to the expected loss, \cite{woods_overstall_2016} used $\tilde{B} = B$ for the Monte Carlo sample size in the inner loop of the DLMC approximation. Again following \cite{woods_overstall_2016}, we fit the Gaussian process model using a squared exponential correlation function. 

The ACE algorithm is implemented in the \texttt{R} package \texttt{acebayes} \citep{acebayes} available from the Comprehensive R Archive Network.

\section{Relationship between proposed approximations and pseudo-Bayesian design} \label{app:pb}

Consider the case where $|M|=1$ and $g(\boldsymbol{\theta}) = \boldsymbol{\theta}$. The approximated SI loss is then
$$\tilde{\lambda}_{SI}(\boldsymbol{\theta},\mathbf{y},\mathbf{d}) = \log \pi(\boldsymbol{\theta}) + \frac{p}{2} \log (2 \pi) + \frac{1}{2} \log | \hat{\Sigma}_{\theta} | + \frac{1}{2}\left(\boldsymbol{\theta} - \hat{\boldsymbol{\theta}}(\mathbf{y})\right)^T\hat{\Sigma}_{\theta}^{-1} \left(\boldsymbol{\theta} - \hat{\boldsymbol{\theta}}(\mathbf{y})\right).$$
The corresponding expected approximate SI loss is given by
\begin{equation}
\tilde{L}_{SI}(\mathbf{d}) = \mathrm{E}_{\boldsymbol{\theta}} \left[ \mathrm{E}_{\mathbf{y}|\boldsymbol{\theta},\mathbf{d}} \left[ \tilde{\lambda}_{SI}(\boldsymbol{\theta},\mathbf{y},\mathbf{d}) \right] \right].
\label{eqn:pbd1}
\end{equation}
Assume that the prior distribution for $\boldsymbol{\theta}$ is sufficiently diffuse so that $\hat{\Sigma}_{\theta} = \mathcal{I}(\boldsymbol{\theta};\mathbf{d})^{-1}$ and the posterior mode is approximately equal to the maximum likelihood estimator (MLE). Using the delta-method and the approximate distribution of the MLE, the following approximation for $\tilde{L}_{SI}(\mathbf{d})$ can be derived
$$\tilde{L}_{SI}(\mathbf{d}) = \mathrm{E}_{\boldsymbol{\theta}} \left[ \frac{p}{2}\log (2 \pi) + \frac{p}{2} + \log \pi(\boldsymbol{\theta}) - \frac{1}{2} \log |\mathcal{I}(\boldsymbol{\theta};\mathbf{d})|\right].$$
This is proportional to the objective function for pseudo-Bayesian D-optimality. Similarly, an approximation to the expected squared error loss is given by
$$\tilde{L}_{SI}(\mathbf{d}) = \mathrm{E}_{\boldsymbol{\theta}} \left[ \mathrm{tr} \left\{ \mathcal{I}(\boldsymbol{\theta};\mathbf{d})^{-1} \right\} \right],$$
the objective function for pseudo-Bayesian A-optimality.

\bibliographystyle{rss}
\bibliography{biblio}

\end{document}